# Sound and Image Processing with Optical Biocomputers

*Cameron L. Jones*

Centre for Mathematical Modelling, School of Mathematical Sciences, Swinburne University of Technology,
Hawthorn VIC 3122, Australia
E-mail: cajones@swin.edu.au

## 1 Introduction

This paper introduces a new method to perform signal processing using cells or atoms to synthesize variation in digital data stored in optical format. Interfacing small-scale biological or chemical systems with information stored on CD, DVD and related disc media has been termed *molecular media*. This is defined as micron or nanoscale interference and/or augmentation of optical data translation with a specific focus on sound and image processing. The principle of this approach uses cells with different morphology and size or atom clusters to geometrically alter the way that binary information is translated into an output signal. These experiments demonstrate that living and non-living, micron-sized cellular systems like bacteria, fungi or yeast [1] and chemical particles, clusters, emulsions or atom arrays [2] having micron or nanosize dimensions can be successfully interfaced with traditional computing devices. This results in multiple hybrid, yet functional information processing devices that can be understood as a *molecular computer* that demonstrates a plurality of useful functions. Comparative studies may be found in [3].

Applications of this method include using molecular scale structures to interfere with data translation during playback of sound, image or video streams for rapid generation of variation. This is particularly important for new media applications, where manipulation of existing data into new arrangements is valuable. The potential market therefore includes graphic designers, filmmakers, music content producers, and all applications that manipulate digital media for design purposes.

CD's and DVD's read digital information that has been encoded onto the surface of the disc as a series of binary bits of specific length that are equivalent to discrete ones and zeros. The laser head reads this string of data and translates this into an output signal (image, sound, text or moving images). To prevent or minimize disc failure due to scratches or smudges on the disc surface, a process called error correction is used to correct these defects. This method exploits the error detection and correction algorithm used during optical disc storage and retrieval, by introducing large numbers of physical errors at a size, frequency and scale that approaches the resolution of the binary bits. In this way, error correction attempts to minimize errors translated at output, yet in practice, the process is corrupted causing fragmentation of data in neighbouring frames of reference to generate nonlinear time series, discontinuous data clustering, and a range of chaotic phenomena.

One can think of the CD or DVD disc to perform the same function as a Petri-plate, where digital content (data) is equivalent to the nutrient medium. Growing cells or creating thin-film chemical deposits in direct contact with the medium establishes a reaction system. The presence of cells or atom clusters alters the medium like a catalyst by changing the rate and/or pathway towards product formation - into sound or images. This simple system is therefore analogous to a bioreactor where cells or enzymes convert reactants into useful products under controlled conditions. As a computational system, this hybrid device is highly sensitive to initial conditions and commonly exhibits power law dependencies that are a hallmark of complex systems [1-2].

This paper is organised as follows: Section 2 reviews nonlinear methods for music composition and how randomness has been used effectively, while new genres like glitch exploit software and/or hardware failure. Section 3 discusses the scale at which contemporary glitch works have been performed, and compares this with glitch generated at the micron and nanoscale. Section 4 details experimental methods for both prokaryotic and eukaryotic cellular systems and recording conditions for both CD-R and DVD-R media, while Section 5 presents examples comparing input with output for both still image and sound. Section 6 examines the concept of molecular computing from an information theory perspective. Section 7 examines some of the relevant mathematics surrounding this version of glitch and illustrates how this relates to complex systems, self-organization, emergence and chaotic dynamics. Section 8 summarises the main results and highlights opportunities for further development.

## 2 Historical Overview of Nonlinear Composition

The generation of new electronic musical scores composed from existing material dates back to pioneering tape-cut-up techniques by John Cage in the 1950's. The principle of tape splicing was to create new timing patterns by recombining tape sections of different size to create a work that was not preconceived. In fact, Cage called such sound, chance music [4] where each performance was necessarily indeterminate. These ideas were extended in the 1960's and 70's by Karlheinz Stockhausen who used tape editing, sound splitting and re-recording to explore the meaning of sound, and its' relationship in space and time. These experiments collectively gave rise to *musique concrète* that developed further into the 1980's, combined with improvements in electronic music synthesis and computational performance. In the mid-80's the record player was increasingly used for new timing effects, while Christian Marclay introduced abusive manipulation like sanding, breaking and cracking to create his unique turntable sound. More recently DJ's and dance-oriented musicians like Oval, Kid 606, Mouse on Mars, Autechre, and Radiohead have used nail varnish, scotch-tape or paint on audio CD's to emulate the damage and error technique first presented by Marclay. In all cases, damage causes failure or glitch [5]. This is an emergent yet unstable process. Glitch generates a unique complex system that overcomes the linear narrative to create unorthodox yet appealing sound scapes.

In contrast, image processing of text, still image or motion pictures using Marclay-type techniques has not been actively explored in practice nor in the literature. To date, glitch manipulation of pixel information has only relied on computer software/hardware failures, DSP at extreme digital resolution or incorrect file format conversions between image or sound files to introduce errors that are then captured or sampled. Glitch art is a new, exciting and emerging field, yet is compromised because unlike digital sound that can be played (post-digital) on many different platforms, image data recorded optically is normally manipulated using computer software, and software is engineered to minimize error, thereby limiting the ability to generate work repeatedly.

## 3 Glitch, Resolution and Pattern Formation

Scratching, breaking and painting CD's introduce gross physical defects at millimetre or more resolution. In contrast, this paper demonstrates various methods to reduce the scale at which physical errors are introduced that span the range from millimetre, through to micron and sub-micron, nanosized errors. Previous experiments with chemical deposits like titanium dioxide, carbon black, aluminium zirconium tetrachlorohydrex, and polyaniline dispersions [1-2] applied as thin films to optical disc media reveal surface coverage of fine-particles and colloidal agglomerates to self-organize into fractal size-shape distribution patterns. This means the laser now traverses the spiral tracking on disc after being filtered through a physical fractal pattern like that shown in Figure 1-3. Laser light attenuation using fractals is a key concept underpinning this application of molecular computing.

It is notable that in the main, fungal cells grow by branching while bacterial and yeast cells multiply by division. This introduces two divergent ways to increase cell size and number, while also changing the density characteristics of surface cover. This no doubt contributes to discontinuities in laser light scattering during processing, since successful reading of data is a form of COPY computation.

## 4 Experimental Methods

An information processing device using bacterial or fungal cells: *Lactobacillus casei* Shirota strain (*Yakult* drink) or *Pycnoporus cinnabarinus* have been grown or applied to the surface of optical media such as CD or DVD discs to cause non-equilibrium error correction during data translation into images or sounds.

*4.1 Commercial DVD and CD Audio*

- Bela Lugosi – Scared to Death / Invisible Ghost, (1941-1946); Flashback Home Entertainment. [DVD]
- Alfred Hitchcock – The 39 Steps, (1935); MRA Entertainment. [DVD]
- Red Hot Chilli Peppers – Blood, Sugar, Sex, Magik; Warner Bros. (1991). [CD-Audio].

*4.2 Eukaryotic Microorganisms*

The filamentous white-rot fungus *Pycnoporus cinnabarinus* that secretes laccase [6-7] was cultivated on Malt Extract Agar, MEA (Oxoid). Spores of *P. cinnabarinus* were spray inoculated [8] onto the data surface of optical discs that had been placed in the base of Petri plates that had been covered with a thin layer of half strength Malt Extract Broth, MEB (Oxoid). The Petri plates were incubated under sterile, humidified conditions at 37ºC for 24 hrs. During growth, nutrient medium evaporation resulted in small, dense colonies forming on the disc surface. Morphological development was similar to growth on solid medium in Petri plate culture. In another version of the experiment, neat aliquots of mycelium could be transferred from the base of the tilted Petri-plate via pipette directly onto the surface of commercial DVD discs. These were incubated at 37ºC for 6 hours to evaporate as much of the residual MEB prior to use, and to allow the fungus to colonise the disc surface to minimize the possibility of dislodgment when the disc is played. Images from motion video were captured from the Alfred Hitchcock DVD, while images from the Bela Lugosi DVD were captured after incubation for 36 hours and then heat treated in an oven at 60° for 2 hrs followed by UV sterilisation for 30min. Note that fungus on discs after 6 hrs was live, while after 36hrs + heating + UV the organism was dead. For the CD-Audio, aliquots were transferred via pipette to the disc surface and incubated for 36 hrs and used immediately. After this time there was significant evaporation, and mycelial clumping was clearly visible. Figure 1 shows cultivation of fungal mycelium directly onto DVD. Scale bar is 100µm.

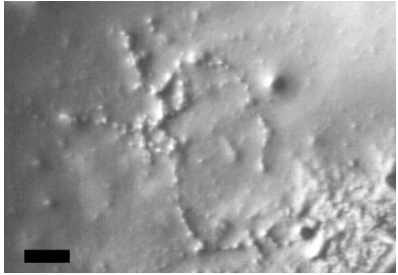

FIGURE 1. Mycelium of *P.cinnabarinus* growing directly on the surface of DVD media after heat fixation.

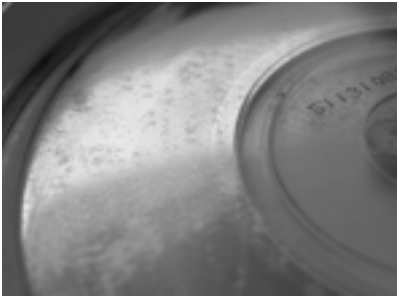

FIGURE 2. Colonies of *P.cinnabarinus* (24hrs) growing on 8cm CD-R (audio) in liquid MEB in a Petri plate.

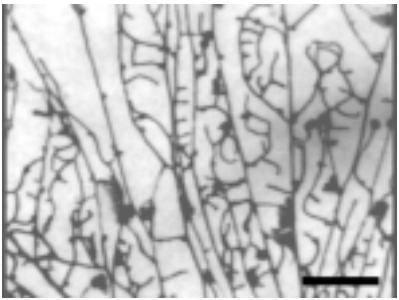

FIGURE 3. Magnification of *P.cinnabarinus* germinated on microporus membranes that have been stained to reveal the branching hyphal network used here as a filter. Scale bar = 100µm.

*4.3 Prokaryotic Microorganisms*

Live cultures of *Lactobacillus casei* Shirota strain from the fermented milk drink *Yakult*, were transferred via streaking directly onto the surface of CD-Audio and evaporated at room temperature until visibly dry. VCD, CD-audio, CD-R, DVD and DVD-R discs were treated in a similar manner. High-resolution micrographs of this organism are shown in [9].

*4.4 CD-R Recording of Audio*

Commercial and originally recorded audio from CD-Audio was WAV encoded (automatic sample rate) in stereo using i-Tunes software on an Apple Macintosh i-Book running OS X (10.2.1) or on the PC (Windows 98 or XP). Playlists of audio were re-recorded to CD-R as WAV files using i-Tunes for the Macintosh or Easy CD Creator for PC. CD-R media was on any of the following: Verbatim DataLifePlus 700MB, TDK CD-R Gold 700MB, or Techworks CD-R 700MB discs. This range included discs that differed in cost and manufacture quality. Alternatively, 8cm round CD-R discs fit well into standard plastic Petri plates (85mm diam.) and were useful for surface incubation of fungi or bacteria.

*4.5 VCD-Recording of Images*

Images in various resolutions and file formats were assembled in Pinnacle Express (Pinnacle Systems) *ver* 1.0.3. A 'black' empty background was selected for central image placement and each image was written to video-CD (VCD) in PAL format. Scene options included a 5 sec length of time for display of each still image. Each VCD could be played in DVD set-top players like the Pioneer DV-344, capable of reading VCD's.

*4.6 DVD-R Recording of Images and/or Sounds*

Alternatively still images, motion video and sound could be recorded to DVD-R using a Powerbook G4 running OS X (10.2.1). i-Movie software was used to assemble data files with a 5 sec slideshow before being exported for assembly directly into i-DVD software. i-DVD recorded to PAL formatted Apple DVD-R media (4.7GB) and several different image sequences could be recorded in this manner.

*4.7 Audio Processing and Capture*

CD-Audio or CD-R discs were played on either: (i) a commercial system consisting of a Gemini CD-9500 ProII professional dual CD player controlled via Roland DJ2000 mixer connected to a Yamaha Natural Sound Power amplifier P-2200 modulated through a McLelland CP-102 dual channel compressor; or (ii) on a Pioneer CDJ1000 CD player or a Pioneer CMX3000 dual CD player or on a portable Walkman style CD player (Avanti). Set-up (ii) was connected to a Behringer VMX300 Pro mixer, connected to a Pioneer VSX-D209 digital signal processor (amplifier). The balanced stereo signal was captured in MP2 format on the Korg Pandora PXR4 digital recording studio or to MiniDisk (Sony MD Walkman, MZ-R37). Minidisk data was transferred to computer using an Edirol USB UA-1A audio capture device. Korg files were transferred to i-Tunes software using a compact flash/smart media device (SanDisk Image Mate). The transferred files were in WAV or AIFF file format at automatic sample rate. Files could be assembled into new playlists and recorded back to CD-R using i-Tunes software. Frequency analysis could be performed using either Cubasis VST, *ver 4* for Macintosh or SoundForge XP *ver 4.5* for PC. Audio was Microsoft RIIF Wave, PCM format, 44,100Hz, 16-bit, stereo signals. To accurately measure differences between disc audio following cell treatments, all WAV data was compared against a clean, control original for each track.

*4.8 Image Processing and Capture*

Images were captured using a Panasonic NV-MX7A digital video camera. Still images could be captured to Multimedia cards from the DV by selecting still photo mode. Images were transferred to either an i-Book or a PowerBook G4 using a Microtech ZiO Multimedia card reader. i-Photo software facilitated this task.

## 5  Image and Audio Results & Discussion

Each computation was begun by placing one disc into the CD or DVD player and pressing play. In most cases the disc was recognized and began playing. If the player failed to recognize the disc, it was necessary to remove some cell material using soft cloth and re-playing until a signal is seen or heard. Note that cells were completely removed from the innermost region of data at the beginning of the disc, since this is header information necessary for the player to mount the disc. It was found that attenuated discs did not play in any of the CD, CD-RW or DVD drives installed in laptop or desktop machines, but only played in stand-alone players. Further tests are needed to clarify this.

Figure 4 shows greyscale versions of original colour images in the first line and treated results following molecular computing in line 2. The still frames show snapshots of Union, Intersection, Negation and fuzzy mutation operations that act by re-combination between adjacent frames of reference.

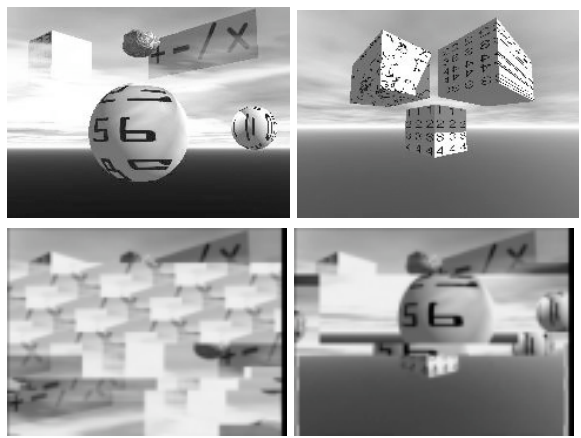

FIGURE 4. Top Row: two original input images; Bottom Row: Still image frame results after molecular computing with *P. cinnabarinus*.

The diversity of abstract patterns generated by this system is a unique feature of cellular processing of digital information, and is a fascinating process to hear or observe. From an applications viewpoint, being able to rapidly generate variation in existing content is a useful adjunct to the creative process.

Notably, this technique has been applied to create graphic design imagery for commercial purposes such as corporate brochure design and for publicity handouts (Swinburne marketing brochures and Blue Velvet Bar pass designs). Figure 5 shows an example of what happens to a fungal-enhanced DVD when played on television. Similar results but with different patterns were seen for *Yakult* treated discs.

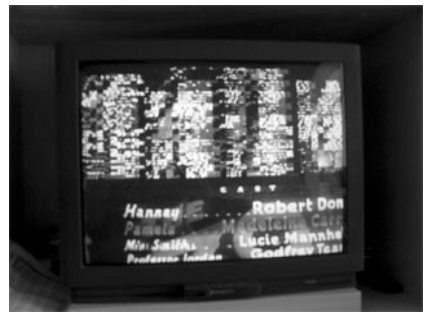

FIGURE 5. Colonies of *P.cinnabarinus* have been cultured onto a DVD disc (Alfred Hitchcock) and played in a Pioneer DVD player through a colour television.

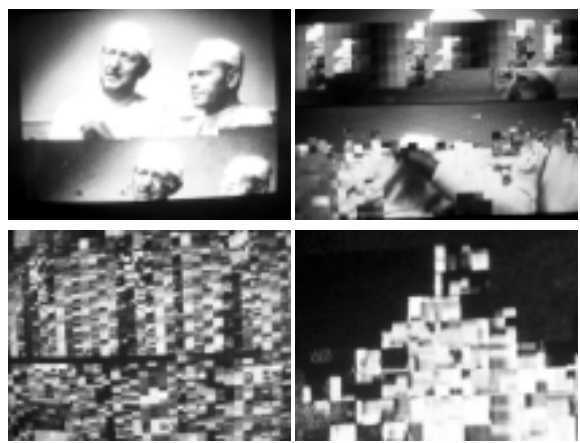

FIGURE 6. Series of single still image screen captures showing results for *P. cinnabarinus* cultured on DVD.

Audio is more difficult to visualise in print publication. New sound arrangements are generated in a similar way to those shown for image information and show changes in pitch, tracking, sample rate and may also be understood in terms of Boolean formulas. Two examples are shown below.

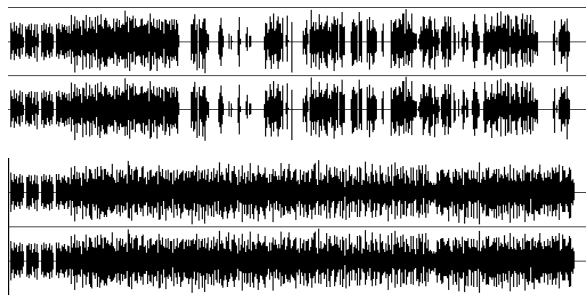

FIGURE 7. Top panel shows the effect of *P. cinnabarinus* on stereo CD-audio for the Red Hot Chilli Peppers (If You Had To Ask), first 90sec played on a Pioneer CMX 3000. The bottom panel shows the untreated control audio.

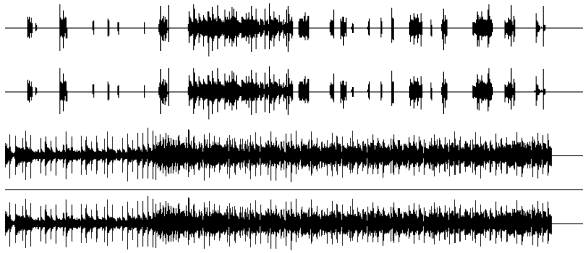

Figure 8. Top panel shows the effect of *P. cinnabarinus* on stereo CD-audio for Red Hot Chilli Peppers (Give It Away), first 30sec played on a portable CD player compared with the untreated control in the bottom row.

Boolean events [10] like <u>copy</u> through AND, OR, NOT, XOR, NAND may be identified in audio output versus the original signal. Suffice to say that new sound arrangements are generated in response to biomolecular filtration of digital data to achieve frequency modulation.

## 6 Molecular Computing & Information Theory

Molecular computing has developed significantly since Richard Feynman discussed life at the very bottom with respect to how small can a number or a computer be? [11]. The crux of molecular computing is its synergy with quantum phenomena [3, 12]. The scale at which events take place is also important for understanding the microphysics of atomic interactions. At very small scales there is often massive parallelism and duality issues that enfold when observations are taken. Control over growth of cells (population dynamics, fractal shape complexity) act as RULES during data translation AXIOMS. In practice any device that shows combinational circuits for AND and NOT can be used to perform any switching function. Image and audio results also show FANOUT and EXCHANGE behaviours further supporting the COPY computational role [12] of this type of molecular device.

## 7 Mathematics of Glitch Phenomena

The principle of error correction is to minimize error! If molecular computing with cells and atoms generates error, then one must consider how many different and unique output sequences (data frames) are generated after say, 8 induced error correction operations (i•ec). When data is amplified or copied under attenuated or induced conditions, this process is compromised and the resulting output is a function of initial conditions and a constant i that defines the complexity of the attenuating field. The result of error correction is a Boolean process with two outcomes, correct or incorrect (1 or 0). If $n_{i•ec}=8$, then there are 256 possible combinations, where the number of distinct i•ec operations for any computation is $2^n$. In practice, the number of burst or block rate errors is much greater than this. Error correction produces a truth table showing where data is stored in optical format. This discretization mechanism (a binary tree) is a strategy for describing a decision sequence for data retrieval. Therefore, image or sound expressions are sum-of-products. In other words, a scene or audio frame-set is a term of a series, and is a product.

Now I will expand on the notational convention [3] originally used for understanding Adleman's DNA computation [13], and show how this helps us trace the behaviour of the four archetype operations that apply to binary data manipulated with cells or atoms. There are four main steps that confer computational ability to molecular systems of the type developed here: (1) merge, (2) amplify, (3) detect, and (4) separate. A simple example is formalised below:

1. input (N)
2. amplify (N) to produce $N_1, N_2, \ldots, N_i$.
3. $Na \leftarrow +(N_1, a)$
4. $Nb \leftarrow +(N_2, b)$
5. $N'b \leftarrow -(Nb, a)$
6. merge (Na, N'b)

What this means is that data is input at step 1, this is amplified by the processor combined with i•ec to produce versions $N_1$, $N_2$ and more. Now using the binary tree look-up-table, extract all versions that express a or b when error correction is satisfied (see function of *w* below). Also produce the set N'b not containing a. Merge Na and N'b. Digital media operated on by molecular methods like the simple model above may therefore be considered to be a multiset of finite objects over a binary alphabet {0,1}. Like in Adleman's experiment with DNA strands, we can consider data frames created and evaluated during error correction to define the following operations:

**Merge:** given media set $N_1$ and $N_2$, the multiset is the union $N_1 \cup N_2$.

**Amplify:** given media set N, produce two copies or more, $N_1$ and $N_2$.

**Detect:** given media set N, return TRUE if N contains at least one correlated media frame, otherwise return FALSE.

**Separate/Extract:** given media set N and a correlated data frame having value *w*, produce two or more copies $+(N,w)$ and $-(N,w)$. Here $+(N,w)$ returns all data frames in N containing a correlated data frame *w*, while $-(N,w)$ returns all data frames in N not having *w* correlations. This generates glitch as error correction and is an iterative procedure. Note, the value of a correlated data frame *w* is dependent on error correction output under attenuated conditions. Image processing of output images could then be used to quantify length and position statistics.

Induced error correction may be considered a type of evolutionary process involving copy computations. For example, a simple crossover computation might combine data frames from adjacent supersets to produce a new image or audio result. Apart from copy computations, results also demonstrate crossover, extinction and mutation events. If one assumes that a piece of data consists of a chunk of 'story', then it is logical to assume that meaning is relational. For sound or image data in optical format, consistency arises over time because of player reliability. Error correction algorithms however create a natural partition to digital data integrity whilst highlighting connection correlations between related frames of reference.

Complexity, nanoscience and self-organization are several interrelated concepts that go some way towards explaining glitch using an experimental modelling approach.

## 8 Conclusions

Surface colonization is a biotechnique similar to microlithography, and in this context, control over growth (population dynamics, fractal shape complexity) act as RULES during data translation AXIOMS. Visual or audio results display a trade-off between strict and weak causality offering micron-level control over synthesised output by manipulating a dynamic complex system.

Differences in cellular parameters at a structural or ultrastructural level including size, organelle type and composition, are considered to be consistent triggers for establishing a critical system that emerges through disruption to error correction using for example, the Cross Interleaved Reed--Solomon Code or variations thereof in optical disc media.

At a local level, many microscopic cell or chemical patterns can be described by fractals. Pattern complexity and its relationship with the entropy displayed after a data transfer COPY step modulated through i•ec might be explained using percolation theory where adjacent frames are evaluated for consistency against weighted nodes available for information flow. Manipulation of digital media through exploiting sensitive dependence on initial conditions is one practical way to control variation. Future experiments will suggest other cultivation strategies for other classes of microorganisms and refine terminology and principles.

Since fungal cells self-organize into fractal shapes [6] it is likely that specific values for percolation in 2d projections should yield an efficient randomisation algorithm. If images or sounds are fragmented into smaller sets of length, L, then the entropy limit is $L\rightarrow\infty$. This property may make this hybrid optical computer of use in cryptoanalysis [14].

Molecular computing with cells and atoms introduces a physical bias towards directed sampling, while nonlinear feedback during error correction look-up-table analysis introduces time displacement.

Molecular computing as described here offers a simple experimental system to deduce cellular automaton rules since any given pattern (sound or image) can be positionally analysed to see where each element of the rule was used [14]. It is notable that cells or atoms act as generating functions.

Universality is often claimed for systems that emulate logic functions like NAND. It is likely that improvements in optical disc storage capacity should accelerate widespread development of quantum algorithms to take advantage of such nano-biomolecular Universal Turing machines.

Examples have been given that highlight the potential of this system for rapid, real-time synthesis and re-synthesis of image or sound streams for multimedia applications or studies of complexity.

## 9 Acknowledgements

This work has been funded through the Chancellery Strategic Initiative and the author would like to particularly thank Prof. Iain Wallace and Prof. Peter Jones for encouragement and assistance. Application of brewers yeast cells directly to CD-audio for micro-glitch was performed first at Blue Velvet Bar & Nightclub, Melbourne, Australia. The author therefore thanks Elisabeth Jones and nanosounds.com for this important milestone and use of equipment.